# Five-dimensional electromagnetism

By D. H. Delphenich (†)
Spring Valley, OH USA

**Abstract:** The usual pre-metric Maxwell equations of electromagnetism in Minkowski space are extended to five dimensions, and the equations of linear five-dimensional electromagnetic waves are obtained from them by using the simplest electromagnetic constitutive law. The role of the dispersion law for five-dimensional plane waves in dictating the most appropriate extension of the four-dimensional Minkowski space metric to the five-dimensional one is discussed. The fact that the five-dimensional dispersion law includes both massless waves and massive ones is emphasized, as well as the fact that one can obtain the Klein-Gordon equation or the Proca equation by separating the fifth coordinate.

(†) E-mail: feedback@neo-classical-physics.info, Website: neo-classical-physics.info

# Contents

**1. Introduction.** – The extension of the space-time manifold from four dimensions to five has been employed for various purposes through the years. Sometimes, as in the case of de Sitter and anti-de-Sitter space-times, as well as Randall-Sundrum models, it is to model the space-time manifold as a submanifold of a higher-dimensional space. One particularly intriguing variation on that theme was the attempts of Vranceanu, Synge, and others [**1**, **2**] to consider that the codimension-one differential system on the five-dimensional manifold that one started with was not actually integrable into a codimension-one foliation of the five-dimensional manifold. Hence, the four-dimensional linear subspaces of the five-dimensional tangent spaces would not take the form of tangent spaces to four-dimensional manifolds, but would still behave like them locally.

By far, the extension to five dimensions that attracted the most attention, and even to the present day, was inaugurated by Theodor Kaluza in 1921 [**3**] and expanded upon in 1926 by Oskar Klein [**4**]. Vladimir Fock [**5**] also deduced essentially the same results as Klein. Kaluza was mostly interested in unifying Einstein's theory of gravitation with Maxwell's theory of electromagnetism by including the components of the electromagnetic potential in the components of the metric on the five-dimensional manifold. Klein, by contrast, was also interested in relativistic wave equations, and in fact, it was in his paper on five-dimensional space-time that he also introduced the Klein-Gordon equation. One obtains the latter equation from the five-dimensional linear wave equation by separating the fifth coordinate from the usual four in a manner that is directly analogous to the way that one goes from the four-dimensional linear wave equation to the Helmholtz equation in wave optics.

Due to the diverging paths that followed from the same basic concept of a five-dimensional extension of the theory of relativity, the Kaluza-Klein program led to further research in the two directions that did not always overlap. For instance, the use of the Kaluza-Klein extension of relativity led to some very deep studies [**6-9**] into the possibility that the extension from four dimensions to five was based in the difference between homogeneous and inhomogeneous coordinates for a four-dimensional projective space. That program came to be called "projective relativity." Pauli [**10**] had even looked into how one could deal with the Dirac equation within that framework. There were also studies that mostly concentrated upon the extension of wave mechanics more than the Einstein-Maxwell unification problem. One might confer the early work by de Broglie [**11**] (see also Klein's comment on that article [**12**]) and Léon Rosenfeld [**13**], as well as the later work by Yurii Rumer [**14**].

The present work is directed more along the latter lines, since the author feels that the experimental verification of "gravito-electromagnetism" had radically altered the theoretical relationship between Einstein's equations and the Maxwell equations by showing that the latter equations are also suitable as a weak-field theory of gravitation, while Einstein's equations describe a strong-field theory. Hence, one should consider the possibility that the two theories are not "ready" to be unified ($^1$). However, the use of five dimensions as a path to a deeper understanding of quantum wave mechanics still remains a valid direction to pursue.

The approach of this article will be to start by naively extending the usual Maxwell equations to five dimensions and examining the extra geometrical objects that are introduced in that way and the mathematical relationships that must follow from imposing the analogous conditions on the five-dimensional fields. Of particular significance is that the "pre-metric" approach to electromagnetism [**16**, **17**] is always kept in at least the background of the discussion. Thus, the Lorentzian metric on space-time will be seen as an artifact of the dispersion law for the propagation of electromagnetic waves, and especially plane waves. That implies that ultimately the fundamental space-time object that drives the theory will be the electromagnetic constitutive law. However, since the extension of four-dimensional constitutive laws to five dimensions is still rather speculative at this point, the only one that will be discussed for heuristic purposes is the one that is usually defined by presence of the Lorentzian metric (i.e., raising the indices).

Eventually, it will be shown how the linear wave equation for five-dimensional electromagnetic waves still follows naturally from Maxwell's equations, just as it did in four dimensions, and that it implies the Klein-Gordon equations by means of separation of variables. The associated dispersion law for (hyper-) plane waves in five dimensions is shown to include the dispersion laws for both massless waves (i.e., photons) and massive ones (i.e., matter waves), without the need to give mass to photons in order to include matter waves. That suggests that the interpretation of the matter waves of

---

($^1$) The author has written a philosophical essay "On the risk of premature unification" [**15**] in which he expands upon that thesis.

quantum mechanics as being basically massive electromagnetic waves that obey the Proca equation, instead of probability waves that obey the Schrödinger equation or its less-used relativistic form, the Klein-Gordon equation, takes on a new plausibility.

The second section of this article will address the five-dimensional extension of the first Maxwell equation, which relates to the electromagnetic field strength 2-form and does not require the use of a metric. In the third section, the extension of the second Maxwell equation, which involves the electromagnetic excitation bivector field, to five dimensions will be discussed, along with the role of the constitutive law. In the following section, the existence and character of wave-like solutions to the Maxwell equations will be examined, along with some of the various dispersion laws that can result, and the extension of such things to five dimensions. In the fifth section, the dispersion laws for quantum wave equations (in particular, Klein-Gordon) will be used as a justification for the extension of the electromagnetic dispersion law to five dimensions, along with some discussion of the five-dimensional extension of the Dirac equation. In the sixth section, the physical interpretation of the fifth dimension will be summarized by looking at how the various fifth components of the physical fields relate to each other. Finally, the results will be summarized, along with suggested topics for deeper analysis.

**2. The first Maxwell equation.** – In order to deal with the first Maxwell equation in five-dimensional form, we shall first introduce the five-dimensional electromagnetic field strength 2-form $^5F$ naively and then show that it can be decomposed with respect to a 4+1 splitting of five-dimensional space-time into the sum of the usual four-dimensional 2-form $F$ and another decomposable 2-form $f \wedge dx^4$, where $f$ is a space-time 1-form. We will then examine how the five-dimensional analogue of the first Maxwell equation decomposes, as well, along with its solution by an electromagnetic potential 1-form.

Although it is becoming traditional to always start with topologically-general differentiable manifolds to represent space-time, we shall start with vector spaces as our manifolds and treat the topological differences between them and more general manifolds as an advanced topic for later discussion. Furthermore, we shall generally treat metrics whose components vary from point to point as another advanced topic that relates to the presence of strong gravitational fields, so the introduction of covariant derivatives will not be necessary.

*a. Notations and conventions.* – A coordinate chart on our five-dimensional vector space $\mathbb{R}^5$ will be written $(x^0, x^1, x^2, x^3, x^4)$, in which the first coordinate is:

$$x^0 = ct. \tag{2.1}$$

The added dimension will then take the form of the final coordinate $x^4$. Hopefully, the structure of the Maxwell equations will eventually give us some insight into the physical nature of $x^4$.

Five-dimensional coordinates and components will employ upper-case Latin symbols and their values will range from 0 to 4. Four-dimensional space-time coordinates will employ lower-case Greek symbols, and their values will range from 0 to 3. Three-dimensional spatial coordinates and components will be expressed by lower-case Latin symbols, while their values will range from 1 to 3.

The natural frame field that is defined by a five-dimensional coordinate system on $\mathbb{R}^5$ will be denoted by $\{\partial_A, A = 0, \ldots, 4\}$, where the notation $\partial_A$ is short for the partial derivative $\partial / \partial x_A$. Its reciprocal coframe field will then be defined by the coordinate differentials $\{dx^A, A = 0, \ldots, 4\}$.

A 4+1 decomposition of $\mathbb{R}^5 = \mathbb{R}^4 \times \mathbb{R}$ will have adapted coordinates that look like $(x^\mu, x^4)$, and a 1 + 3 decomposition of $\mathbb{R}^4 = \mathbb{R} \times \mathbb{R}^3$ will have adapted coordinates that look like $(x^0, x^i)$.

Since it is not necessary to introduce a metric on space-time in order to define the first set of Maxwell equations, we shall defer a discussion of that issue until later.

*b. The electromagnetic field strength 2-form.* – Naively, the five-dimensional electromagnetic field strength 2-form can be expressed locally in the form:

$$^5F = \tfrac{1}{2} F_{AB}\, dx^A \wedge dx^B . \qquad (2.2)$$

Note that all of the components $F_{AB}$ are assumed to be functions of all five coordinates.

If we perform a 4+1 decomposition of $\mathbb{R}^5$ then $^5F$ will take the form:

$$^5F = F + f \wedge dx^4, \qquad (2.3)$$

in which we have defined:

$$F = \tfrac{1}{2} F_{\mu\nu}\, dx^\mu \wedge dx^\nu, \qquad f = F_{\mu 4}\, dx^\mu. \qquad (2.4)$$

Hence, the space-time 2-form $F$ takes the usual form that it has in the formulation of Maxwell's equations on Minkowski space, except that its components are functions of five coordinates, not four. The space-time 1-form $f$ is the main novelty that has been introduced by going to the five-dimensional picture.

When one performs a 1+3 splitting (i.e., time plus space) on $\mathbb{R}^4$, $F$ will take the usual form:

$$F = c\, dt \wedge E - B, \qquad (2.5)$$

in which:

$$E = F_{0i}\, dx^i, \qquad B = \tfrac{1}{2} B_{ij}\, dx^i\, dx^j \qquad (2.6)$$

are the spatial electric field strength 1-form and the spatial magnetic flux density 2-form, respectively.

Meanwhile, the 1-form $f$ will split into

$$f = c f_0\, dt + f_s, \qquad f_s = f_i\, dx^i. \qquad (2.7)$$

*d. The Lorentz force.* – In four-dimensional electromagnetism, one gets from the electromagnetic field strength 2-form $F$ to the force $\mathfrak{f}$ that it exerts upon an electric current $\mathbf{J}$ that flows in the field $F$ by a simple interior product:

$$\mathfrak{f} = i_{\mathbf{J}} F, \tag{2.8}$$

which locally looks like:

$$\mathfrak{f}_\mu = F_{\mu\nu} J^\nu. \tag{2.9}$$

When $F$ has the 1+3 form in (2.5) and $\mathbf{J} = \rho_e \, (\frac{1}{c} \partial_t + \mathbf{v})$, $\mathfrak{f}$ will be:

$$\mathfrak{f} = -c\, \rho_e\, E(\mathbf{v})\, dt + \rho_e\, (E + i_{\mathbf{v}} B), \tag{2.10}$$

the spatial part of which will agree with the usual Lorentz force when one expresses the spatial 2-form $B$ as the spatial dual $\#_s \mathbf{B}$ of the magnetic flux density vector field $\mathbf{B}$:

$$B = \#_s \mathbf{B} = i_{\mathbf{B}} V_s, \qquad V_s = dx^1 \wedge dx^2 \wedge dx^3 = \frac{1}{3!} \varepsilon_{ijk}\, dx^i \wedge dx^j \wedge dx^k, \tag{2.11}$$

which will make the components of $i_{\mathbf{v}} B$ take the form:

$$(i_{\mathbf{v}} B)_i = \varepsilon_{ijk}\, v^j B^k = (\mathbf{v} \times \mathbf{B})_i \, ; \tag{2.12}$$

i.e., the usual cross product.

The obvious five-dimensional extension of (2.8) would be:

$$^5\mathfrak{f} = i_{^5\mathbf{J}} \, ^5F. \tag{2.13}$$

With $^5F$ in the form (2.3) and:

$$^5\mathbf{J} = \mathbf{J} + J^4 \partial_4, \tag{2.14}$$

$^5\mathfrak{f}$ will become:

$$^5\mathfrak{f} = \mathfrak{f} - J^4 f + f(\mathbf{J})\, dx^4. \tag{2.15}$$

Hence, the usual Lorentz force $\mathfrak{f}$ has picked up a space-time contribution that is proportional to $f$ and a component $f(\mathbf{J})$ that shows up in the added fifth dimension.

*d. The first Maxwell equation.* – The obvious five-dimensional extension of the first Maxwell equation in space-time:

$$d_\wedge F = 0 \tag{2.16}$$

is

$$^5d_\wedge \, ^5F = 0. \tag{2.17}$$

In this, we are using the notation $^5d_\wedge$ in order to denote the five-dimensional extension of the exterior derivative operator on space-time, which is denoted by $d_\wedge$. The main difference is that one has to consider the partial derivative with respect to $x^4$, along with the others. Hence, with $^5F$ represented in the form (2.3), one will have:

$$^5d_\wedge {}^5F = {}^5d_\wedge F + {}^5d_\wedge(f \wedge dx^4) = d_\wedge F + (\partial_4 F + d_\wedge f) \wedge dx^4, \quad (2.18)$$

in which:

$$d_\wedge F = \tfrac{1}{3}(\partial_\kappa F_{\mu\nu} + \partial_\mu F_{\nu\kappa} + \partial_\nu F_{\kappa\mu}) dx^\kappa \wedge dx^\mu \wedge dx^\nu \quad (2.19)$$

is the usual space-time expression, while:

$$\partial_4 F = \tfrac{1}{2}(\partial_4 F_{\mu\nu}) dx^\mu \wedge dx^\nu, \qquad d_\wedge f = \tfrac{1}{2}(\partial_\mu f_\nu - \partial_\mu f_\nu) dx^\mu \wedge dx^\nu. \quad (2.20)$$

Hence, since the two terms in the right-hand side of (2.18) are linearly independent, (2.17) will split into the pair of equations:

$$d_\wedge F = 0, \qquad \partial_4 F = - d_\wedge f. \quad (2.21)$$

(We have also used the fact that $dx^4$ is linearly independent of all of the space-time covectors.)

The first of these equations is the usual Maxwell equation for space-time. The second one is more specific to the added dimension and takes the form of an "induction" equation that couples the partial derivative of the electromagnetic field with respect to the fifth coordinate to the exterior derivative of the somewhat-enigmatic space-time 1-form $f$.

*e. The electromagnetic potential 1-form.* – The Poincaré lemma is just as valid for five dimensions as it was for four dimensions, so equation (2.17) implies that there will be a global (since we are dealing with a vector space) five-dimensional 1-form $^5A$ such that:

$$^5F = {}^5d_\wedge {}^5A. \quad (2.22)$$

A 4+1 splitting will make:

$$^5A = A + A_4\, dx^4, \quad (2.23)$$

in which $A = A_\mu\, dx^\mu$ is the usual space-time 1-form that one introduces, except that its components are functions of five coordinates, not four, now. One will then have:

$$^5d_\wedge {}^5A = d_\wedge A + (dA_4 - \partial_4 A) \wedge dx^4, \quad (2.24)$$

and with consideration given to (2.3), as well as linear independence, (2.22) will split into the pair of equations:

$$F = d_\wedge A, \qquad f = dA_4 - \partial_4 A. \quad (2.25)$$

Hence, we have recovered the usual space-time relationship in the first of these equations, while the second one casts a different light upon the nature of the fifth-dimensional contributions. In component form, the latter equations are:

$$F_{\mu\nu} = \partial_\mu A_\nu - \partial_\nu A_\mu, \qquad f_\mu = \partial_\mu A_4 - \partial_4 A_\mu. \quad (2.26)$$

The concept of gauge invariance of the five-dimensional 2-form $^5F$ is directly analogous to the four-dimensional version: Two potential 1-forms $^5A$ and $^5A'$ will produce the same field strength 2-form $^5F$ iff $^5A' - {}^5A$ is a closed 1-form. Since we are dealing with a vector space (hence, simply-connected), the closed 1-form can be represented by an exact 1-form $^5d\, ^5\lambda$, where $^5\lambda$ is a 0-form (i.e., smooth function) on $\mathbb{R}^5$. Hence, a gauge transformation (of the second kind) of $^5A$ would replace it with $^5A + {}^5d\, ^5\lambda$.

**3. The second Maxwell equation.** – The second Maxwell equation (div $\mathfrak{H} = -4\pi \mathbf{J}$) requires a bit more preliminary discussion, because in its conventional relativistic formulation, one is basically using the simplest electromagnetic constitutive law to associate the electromagnetic excitation bivector field $\mathfrak{H}$ with the field strength 2-form $F$, namely, the one for the classical electromagnetic vacuum. That takes the form of simply introducing the Lorentzian metric on space-time and using the linear isomorphisms that it defines between tangent and cotangent spaces (more precisely, their inverses) to map 2-forms to bivectors; in components, those linear isomorphisms amount to the "raising of both indices."

A more modern approach to electrodynamics is the so-called "pre-metric" approach that goes back to work of Kottler [18] and van Dantzig [19], as well as a comment by Cartan [20] on the subject ([1]). In that approach, one starts with a more general constitutive law and recovers the Lorentzian metric (i.e., the light cones) as a degenerate case of a more general dispersion law for the propagation of electromagnetic waves.

Another point that is never observed is that since the soul of the divergence operator is rooted in the concept of a volume element, it is more natural to think of the divergence operator as something that acts upon multivector fields, rather than exterior differential forms. Indeed, it takes the form of the adjoint of the exterior derivative operator under the Poincaré isomorphism that a volume element will define.

*a. The electromagnetic excitation bivector field.* – Before we introduce a constitutive law, we can at least define the five-dimensional electromagnetic excitation bivector field $^5\mathfrak{H}$ abstractly and impose the Maxwell equation upon it as an axiom.

The bivector field $^5\mathfrak{H}$ can be written in component form as:

$$^5\mathfrak{H} = \tfrac{1}{2}\, ^5\mathfrak{H}^{AB}\, \partial_A \wedge \partial_B\,. \tag{3.1}$$

Hence, when a 4+1 splitting of $\mathbb{R}^5$ is made, it can be decomposed into a sum:

$$^5\mathfrak{H} = \mathfrak{H} + \mathbf{h} \wedge \partial_4\,, \tag{3.2}$$

---

([1]) For more modern discussions of pre-metric electromagnetism, see the book by Hehl and Obukhov [16] and the author's own book [17], which essentially complements their discussion with expanded treatments of various aspects of the theory.

in which:
$$\mathfrak{H} = \tfrac{1}{2} {}^5\mathfrak{H}^{\mu\nu} \partial_\mu \wedge \partial_\nu \qquad (3.3)$$

is the usual space-time bivector field, and:

$$\mathbf{h} = {}^5\mathfrak{H}^{\mu 4} \partial_\mu \qquad (3.4)$$

is a space-time vector field that is, in a sense, dual to the 1-form *f*. (Once again, remember that all components depend upon five coordinates now.)

When one further performs a 1+3 split on $\mathbb{R}^4$, $\mathfrak{H}$ will decompose into:

$$\mathfrak{H} = c\, dt \wedge \mathbf{D} + \mathbf{H}, \qquad (3.5)$$

in which:
$$\mathbf{D} = \mathfrak{H}^{0i} \partial_i, \qquad \mathbf{H} = \tfrac{1}{2} \mathfrak{H}^{ij} \partial_i \wedge \partial_j \qquad (3.6)$$

are the spatial vector field that represents the electric excitations (often called the "electric displacement") and the magnetic field strength bivector field, which is also a spatial tensor field, respectively.

*b. The divergence operator.* – In order to pose the second Maxwell equation, we need to first explain how one extends the divergence operator from vector fields to multivector fields. More to the point, we need to do that in five dimensions.

First, one introduces a volume element on $\mathbb{R}^5$. (There are no topological issues with orientability in that case.) It will take the form of an everywhere non-zero 5-form ${}^5V$, which can take the local component form:

$$^5V = dx^0 \wedge \ldots \wedge dx^4 = \frac{1}{5!} \varepsilon_{A_0 \cdots A_4} dx^{A_0} \wedge \cdots \wedge dx^{A_4}, \qquad (3.7)$$

in which $\varepsilon_{A_0 \cdots A_4}$ is the five-dimensional Levi-Civita symbol, whose components are completely antisymmetric, and the non-zero components will therefore represent permutations of the sequence 01234, with a sign that is positive for an even permutation and negative for an odd one.

The volume element will allow one to define a series of linear isomorphisms ${}^5\#$ : $\Lambda_k \mathbb{R}^5 \to \Lambda^{5-k} \mathbb{R}^5$, ${}^5\mathbf{B} \mapsto {}^5\# \, {}^5\mathbf{B}$, where the (5–*k*)-form that corresponds to the *k*-vector field ${}^5\mathbf{B}$ is defined by the interior product:

$$^5\# \, {}^5\mathbf{B} \equiv i_{{}^5\mathbf{B}} {}^5V. \qquad (3.8)$$

In components, that takes the form:

$$^5 B_{A_0 \cdots A_{4-k}} = \frac{1}{k!} \varepsilon_{A_0 \cdots A_4} \, {}^5 B^{A_{5-k} \cdots A_4}. \qquad (3.9)$$

Since the linear map $^5\#$ is an isomorphism, it can be inverted, and one way of characterizing the inverse isomorphism is to start with an everywhere-nonzero 5-vector field $^5\mathbf{V}$, which might locally look like:

$$^5\mathbf{V} = \partial_0 \wedge \ldots \wedge \partial_4 = \frac{1}{5!} \varepsilon^{A_0 \cdots A_4} \partial_{A_0} \wedge \cdots \wedge \partial_{A_4}, \tag{3.10}$$

and as a result, this particular choice will have the property that:

$$^5V(^5\mathbf{V}) = \det [\partial_0 \mid \ldots \mid \partial_4] = 1. \tag{3.11}$$

This inverse volume element will define the inverse isomorphisms $^5\#^{-1} : \Lambda^k \mathbb{R}^5 \to \Lambda_{5-k} \mathbb{R}^5$, $^5F \mapsto {^5\#^{-1}}\, ^5F$ for each $k = 0, \ldots, 4$ by an analogous interior product of a dual nature.

Between the isomorphism $^5\#$ and its inverse, one can map the exterior derivative operator $^5d_\wedge$ to an adjoint operator $^5\text{div} : \Lambda_k \mathbb{R}^5 \to \Lambda_{k-1} \mathbb{R}^5$, that takes an $k$-vector field $^5\mathbf{B}$ to:

$$^5\text{div}\, ^5\mathbf{B} = (^5\#^{-1} \cdot {^5d_\wedge} \cdot {^5\#})\, ^5\mathbf{B}. \tag{3.12}$$

This divergence operator then takes $k$-vector fields to $(k-1)$-vector fields and has a property that is analogous to the property of $^5d_\wedge$ that its square will vanish identically:

$$^5\text{div} \cdot {^5\text{div}} = (^5\#^{-1} \cdot {^5d_\wedge} \cdot {^5\#}) \cdot (^5\#^{-1} \cdot {^5d_\wedge} \cdot {^5\#}) = {^5\#^{-1}} \cdot ({^5d_\wedge} \cdot {^5d_\wedge}) \cdot {^5\#} = 0. \tag{3.13}$$

Furthermore, the divergence operator agrees with the conventional one on vector fields. Hence, if $^5\mathbf{X} = X^A \partial_A$ is a five-dimensional vector field then:

$$^5\text{div}\, ^5\mathbf{X} = \partial_A X^A, \tag{3.14}$$

and if $^5\mathbf{B} = \tfrac{1}{2} B^{AB} \partial_A \wedge \partial_B$ is a five-dimensional bivector field then:

$$^5\text{div}\, ^5\mathbf{B} = \partial_A B^{AB}. \tag{3.15}$$

If one performs a 4+1 split on $\mathbb{R}^5$ then the latter equation will decompose into:

$$^5\text{div}\, ^5\mathbf{B} = \text{div}\, \mathbf{B} - \partial_4 \mathbf{b} + (\text{div}\, \mathbf{b})\, \partial_4, \tag{3.16}$$

in which we have defined the space-time bivector field and vector field:

$$\mathbf{B} = \tfrac{1}{2} B^{\mu\nu} \partial_\mu \wedge \partial_\nu, \quad \mathbf{b} = B^{\mu 4} \partial_\mu, \tag{3.17}$$

respectively, and div refers to the space-time divergence operator.

*c. The second Maxwell equation.* – We are now in a position to define the five-dimensional extension of the second Maxwell equation, namely:

$$^5\text{div }{}^5\mathfrak{H} = -4\pi\, {}^5\mathbf{J}, \tag{3.18}$$

in which $^5\mathbf{J}$ is a five-dimensional electric current vector field that serves as the source of the excitations.

Under a 4+1 split, (3.18), with the use of (3.16) and (2.14), will yield:

$$\text{div }\mathfrak{H} - \partial_4 \mathbf{h} = -4\pi\mathbf{J}, \qquad \text{div }\mathbf{h} = -4\pi J^4, \tag{3.19}$$

so $\mathfrak{H}$ will be the usual space-time bivector field of electromagnetic excitations, and $\mathbf{J}$ will be the usual space-time electric current source vector field (although both $\mathfrak{H}$ and $\mathbf{J}$ will still be functions of five coordinates, not four), while $\mathbf{h}$ was defined in (3.4).

The operator div is simply the space-time divergence operator. $\mathbf{h}$ seems to relate to $J^4$ in the same way that $\mathbf{D}$ relates to the electric charge density $J^0 = \rho$, although one should keep in mind that div is a four-dimensional divergence, not a three-dimensional one, and will therefore include a time derivative.

A brief glance at the first equation in (3.19) will show that it differs from the usual relativistic Maxwell equation by the appearance of the term $-\partial_4\mathbf{h}$. One can also move that term to the right-hand side and treat it as a contribution to the source current, such as a polarization current.

*d. Conservation of charge.* – When one takes the divergence of both sides of the second Maxwell equation (3.18), one will find that due to the property of the divergence operator (3.13), one will get the identity:

$$^5\text{div }{}^5\mathbf{J} = 0, \tag{3.20}$$

which generalizes the four-dimensional condition on $\mathbf{J}$ that represents not only the conservation of charge, but a compatibility – or *integrability* – condition on the current vector fields that can be used as sources for the $\mathfrak{H}$ field.

Of course, since we are dealing with five dimensions now, not four, a 4+1 split will turn (3.20) into:

$$\text{div }\mathbf{J} = -\partial_4 J^4. \tag{3.21}$$

Hence, the four-dimensional compatibility condition for the space-time current $\mathbf{J}$ that its divergence must vanish will now have a potentially-nonzero contribution from the fifth dimension.

*e. Electromagnetic constitutive laws.* – Let us assemble the 4+1 forms of both Maxwell equations (2.21), (3.19), along with the identity (3.21):

$$\boxed{\begin{array}{ll} d_\wedge F = 0, & \partial_4 F = -d_\wedge f, \\ \operatorname{div} \mathfrak{H} - \partial_4 \mathbf{h} = -4\pi \mathbf{J}, & \operatorname{div} \mathbf{h} = -4\pi J^4, \quad \operatorname{div} \mathbf{J} = -\partial_4 J^4. \end{array}} \qquad (3.22)$$

The unknowns in these equations are the space-time tensor fields $F$, $f$, **H**, **h**, while **J** and $J^4$ are known. Hence, there are twenty unknown components, in all. Meanwhile, there are fifteen equations (not including the identity). Therefore, the system is underdetermined.

A first step in reducing the number of unknowns is to define an invertible functional relationship between the field strengths, which are treated as stimuli in the electromagnetic medium in question, and the excitations, which then represent the response of the medium to the stimulus. In conventional electromagnetism, the difference between the field strengths and the excitations is defined by the formation of electric dipoles, as in dielectrics, and magnetic dipoles, as in magnetic materials. One refers to that process as the "polarization" of the medium under the presence of imposed electromagnetic fields.

In four dimensions, an electromagnetic constitutive law will then be an invertible, differentiable relationship with a differentiable inverse (i.e., a diffeomorphism):

$$\mathfrak{H} = \mathfrak{H}(F), \qquad (3.23)$$

and when one makes a 1+3 split, one will have a pair of invertible, differentiable relationships:

$$\mathbf{D} = \mathbf{D}(E, \mathbf{H}), \qquad B = B(E, \mathbf{H}). \qquad (3.24)$$

Note that this implies that the usual way of constructing $F$ and $\mathfrak{H}$ seems to be mixing field strengths with excitations, and vice versa.

The obvious extension of (3.23) would be an invertible, differentiable relationship with a differentiable inverse:

$$^5\mathfrak{H} = {}^5\mathfrak{H}(^5F). \qquad (3.25)$$

A 4+1 split would decompose this into two such relationships:

$$\mathfrak{H} = \mathfrak{H}(F, f), \qquad \mathbf{h} = \mathbf{h}(F, f). \qquad (3.26)$$

Hence, in addition to the somewhat-mysterious second relationship, the conventional space-time one (3.23) has picked up a contribution from the fifth dimension, which should presumably cause electric and magnetic dipoles to appear that would have a purely novel character.

In conventional (i.e., not pre-metric) electromagnetism, one typically deals with the simplest possible constitutive law, which would be defined by the classical electromagnetic vacuum. Such a law is linear, isotropic, homogeneous, and non-dispersive, in the sense that the relationship is purely algebraic and does not include any integral operators. One can then set:

$$\mathbf{D} = \varepsilon_0\, \mathbf{E}, \qquad \mathbf{B} = \mu_0\, \mathbf{H}, \qquad (3.27)$$

in which $\varepsilon_0$ is the vacuum dielectric constant, and $\mu_0$ is its magnetic permeability. Actually, these associations also implicitly contain a linear isomorphism that will convert spatial vectors to spatial covectors, such as might be defined by a spatial Euclidian metric. That is, in components, one would usually have:

$$D^i = \varepsilon_0\, \delta^{ij} E_j, \qquad B^i = \mu_0\, \delta^{ij} H_j. \qquad (3.28)$$

Since we have previously defined $B$ to be a spatial 2-form and $\mathbf{H}$ to be a spatial bivector field, the way that one converts them back to the more conventional vector field $\mathbf{B}$ and covector field $H$ is to use the spatial duality isomorphism that is defined by a spatial volume element:

$$V_s = dx^1 \wedge dx^2 \wedge dx^3 = \frac{1}{3!} \varepsilon_{ijk}\, dx^i \wedge dx^j \wedge dx^k. \qquad (3.29)$$

One then defines $\mathbf{B}$ and $H$ by:

$$B = \#_s \mathbf{B} = i_{\mathbf{B}} V_s \quad (B_{ij} = \varepsilon_{ijk} B^k), \qquad H = \#_s \mathbf{H} = i_{\mathbf{H}} V_s \quad (H_i = \tfrac{1}{2} \varepsilon_{ijk} H^{jk}). \qquad (3.30)$$

Indeed, the only way that $\varepsilon_0$ and $\mu_0$ typically show up in the vacuum Maxwell equations is by way of:

$$c = \frac{1}{\sqrt{\varepsilon_0\, \mu_0}}. \qquad (3.31)$$

Since that quantity is often set equal to 1, the role of electromagnetic constitutive laws is generally opaque to a good number of theoretical physicists.

When we have defined a constitutive law of the form (3.25) or (3.26), the number of unknown components will be halved to ten unknowns. Of course, since we still have fifteen equations, that will now make the system overdetermined. In order to make the system well-determined, we need only to solve the first Maxwell equation for a potential 1-form $^5A$, which will have five unknown components. Using (2.25), the equations (3.22) will then take the form:

$$\boxed{\begin{array}{c} F = d_\wedge A, \qquad f = dA_4 - \partial_4 A, \\ \operatorname{div} \mathfrak{H} - \partial_4 \mathbf{h} = -4\pi \mathbf{J}, \quad \operatorname{div} \mathbf{h} = -4\pi J^4, \quad \operatorname{div} \mathbf{J} = -\partial_4 J^4. \end{array}} \qquad (3.32)$$

Since the top row consists of identities, not equations, and the assumption of a constitutive law would make the unknowns in the second row (namely, $\mathfrak{H}$ and $\mathbf{h}$) into functions of the five unknowns $A$, $A_4$ (or really, their partial derivatives), and since there are five equations in the second row, that would make the resulting system of partial differential equations well-determined.

Note that since $\mathfrak{H}$ and $\mathbf{h}$ are always differentiated in the second row of (3.32), it is sufficient to know the partial derivatives:

$$K^{\mu\nu\kappa\lambda} \equiv \frac{\partial \mathfrak{H}^{\mu\nu}}{\partial F_{\kappa\lambda}}, \qquad L^{\mu\nu\kappa} \equiv \frac{\partial \mathfrak{H}^{\mu\nu}}{\partial f_{\kappa}}, \qquad M^{\mu\kappa\lambda} \equiv \frac{\partial h^{\mu}}{\partial F_{\kappa\lambda}}, \qquad N^{\mu\kappa} \equiv \frac{\partial h^{\mu}}{\partial f_{\kappa}}.$$

The constitutive law will be linear iff these derivatives are all functions of only $x^A$. In space-time electromagnetism, only the first set of partial derivatives is required, so the other three are the ones that require further analysis.

Note that certain symmetries will be inherent to the components of the constitutive law:

1. $K^{\mu\nu\kappa\lambda}$ is antisymmetric in $\mu\nu$ and $\kappa\lambda$ individually.
2. $L^{\mu\nu\kappa}$ is antisymmetric in $\mu\nu$.
3. $M^{\mu\kappa\lambda}$ is antisymmetric in $\kappa\lambda$.

Other symmetries are possible. For instance, $K^{\mu\nu\kappa\lambda}$ is often symmetric in the exchange of the pair $\mu\nu$ with the pair $\kappa\lambda$, due to reciprocity.

When one includes these relationships in the second row of (3.32), one will get, in component form:

$$\tfrac{1}{2} K^{\mu\nu\kappa\lambda} \frac{\partial F_{\kappa\lambda}}{\partial x^{\mu}} - \tfrac{1}{2} M^{\nu\kappa\lambda} \frac{\partial F_{\kappa\lambda}}{\partial x^4} + L^{\mu\nu\kappa} \frac{\partial f_{\kappa}}{\partial x^{\mu}} - N^{\nu\kappa} \frac{\partial f_{\kappa}}{\partial x^4} = -4\pi J^{\nu}, \tag{3.33}$$

$$\tfrac{1}{2} M^{\mu\kappa\lambda} \frac{\partial F_{\kappa\lambda}}{\partial x^{\mu}} + N^{\mu\kappa} \frac{\partial f_{\kappa}}{\partial x^{\mu}} = -4\pi J^4. \tag{3.34}$$

When one expresses $F$ and $f$ in terms of $A$, $A_4$, these equations will become:

$$K^{\mu\nu\kappa\lambda} \partial_{\mu\kappa} A_{\lambda} - M^{\nu\kappa\lambda} \partial_{4\kappa} A_{\lambda} + L^{\mu\nu\kappa} (\partial_{\mu\kappa} A_4 - \partial_{\mu 4} A_{\kappa}) - N^{\nu\kappa} (\partial_{4\kappa} A_4 - \partial_{44} A_{\kappa})$$
$$= -4\pi J^{\nu}, \tag{3.35}$$

$$M^{\mu\kappa\lambda} \partial_{\mu\kappa} A_{\lambda} + N^{\mu\kappa} (\partial_{\mu\kappa} A_4 - \partial_{\mu 4} A_{\kappa}) = -4\pi J^4. \tag{3.36}$$

Little more can be said at this point in advance of more detailed knowledge of the constitutive law, except that in four dimensions, one would have only:

$$K^{\mu\nu\kappa\lambda} \partial_{\mu\kappa} A_{\lambda} = -2\pi J^{\nu} \tag{3.37}$$

as the field equations for $A_{\mu}$.

**4. Electromagnetic waves.** – We shall first review the way that one usually gets electromagnetic waves in four dimensions and then look at the process of extending that to five.

*a. The wave equation in four dimensions.* – In four-dimensional electromagnetism, the conventional approach (as opposed to the pre-metric one) to defining the constitutive law that associates the bivector field $\mathfrak{H}$ with the 2-form $F$ is to assume that there is a Lorentzian metric $g$ on space-time that takes the local form:

$$g = g_{\mu\nu}\, dx^{\mu}\, dx^{\nu} \qquad (g_{\mu\nu} = g_{\nu\mu}), \tag{4.1}$$

in which the multiplication of the 1-forms is a symmetrized tensor product.

In an orthonormal (or *Lorentzian*) local coframe field $\{\theta^{\mu}, \mu = 0, \ldots, 3\}$, $g$ will take the form:

$$g = \eta_{\mu\nu}\, \theta^{\mu}\, \theta^{\nu}, \qquad \eta_{\mu\nu} = \text{diag}\,[+1, -1, -1, -1]. \tag{4.2}$$

Whether or not a coordinate system always exists whose natural coframe field $dx^{\mu}$ is Lorentzian is a deep problem in the integrability of "*G*-structures." Let it suffice to say that it has a lot to do with the vanishing of the Riemannian curvature tensor that $g$ defines, so such coordinate charts will exist on flat spaces, such as Minkowski space.

When a Lorentzian metric has been defined, one can also define a linear isomorphism of each tangent space $T_x M$ to the corresponding cotangent space $T_x^* M$ that takes each tangent vector $\mathbf{u} \in T_x M$ to the covector $i_{\mathbf{u}}\, g \in T_x^* M$, which has the defining property that for any tangent vector $\mathbf{v} \in T_x M$, the evaluation of the covector $i_{\mathbf{u}}\, g$ on the vector $\mathbf{v}$ will produce the scalar:

$$i_{\mathbf{u}}\, g\,(\mathbf{v}) = g\,(\mathbf{u}, \mathbf{v}) = g_{\mu\nu}\, u^{\mu}\, v^{\nu}. \tag{4.3}$$

Hence, the components of $i_{\mathbf{u}}\, g$ will take the form:

$$(i_{\mathbf{u}}\, g)_{\nu} = u_{\nu} = g_{\mu\nu}\, u^{\mu}. \tag{4.4}$$

In other words, the linear isomorphism $\iota_g : T_x M \to T_x^* M$ that $g$ defines is "lowering the index." Its inverse $\iota_g^{-1} : T_x^* M \to T_x M$ will therefore be "raising the index":

$$u^{\mu} = g^{\mu\nu}\, u_{\nu}, \qquad g^{\mu\kappa}\, g_{\kappa\nu} = \delta_{\nu}^{\mu}. \tag{4.5}$$

One can use $\iota_g^{-1}$ to map 2-forms in $\Lambda_x^* M = T_x^* M \wedge T_x^* M$ by forming essentially the "exterior product" $\iota_g^{-1} \wedge \iota_g^{-1} : \Lambda_x^* M \to \Lambda_{2,x} M$, which basically raises both indices of a 2-form $F$:

$$F^{\mu\nu} = g^{\mu\kappa}\, g^{\nu\lambda}\, F_{\kappa\lambda} = \tfrac{1}{2}\,(g^{\mu\kappa}\, g^{\nu\lambda} - g^{\mu\lambda}\, g^{\nu\kappa})\, F_{\kappa\lambda}. \tag{4.6}$$

Hence, one can use $\iota_g^{-1} \wedge \iota_g^{-1}$ as the electromagnetic constitutive law that has the components $K^{\mu\nu\kappa\lambda}$ above; i.e.:

$$K^{\mu\nu\kappa\lambda} = g^{\mu\kappa}\, g^{\nu\lambda} - g^{\mu\lambda}\, g^{\nu\kappa}. \tag{4.7}$$

When one substitutes that in the field equation (3.37) for $A_{\lambda}$, one will get:

$$g^{\nu\lambda} (\Box_g A_\lambda - g^{\mu\kappa} \partial_{\mu\lambda} A_\kappa) = -4\pi J^\nu \qquad (\Box_g \equiv g^{\mu\nu} \partial_{\mu\nu}) \qquad (4.8)$$

or

$$\Box_g A_\nu - g^{\mu\kappa} \partial_{\mu\nu} A_\kappa = -4\pi J_\nu, \qquad (4.9)$$

In the event that one is in Minkowski space, so $g^{\mu\kappa} = \eta^{\mu\kappa}$ is constant, the second term on the left will become $\partial_{\mu\nu} A^\nu$ and if one imposes the Lorentz gauge upon the 1-form $A$:

$$0 = \text{div } \mathbf{A} = \partial_\nu A^\nu \qquad (4.10)$$

then what will be left of the field equations for $A$ is:

$$\Box A_\mu = -4\pi J_\mu, \qquad (4.11)$$

which is the linear wave equation with a source term.

*b. Four-dimensional dispersions laws.* – More generally, one must specify the electromagnetic properties of the medium in order to get the constitutive map $K$ when it does not take the metric component form (4.7), and that is the soul of pre-metric electromagnetism. In order to retrieve the actual space-time metric when one starts with the electromagnetic constitutive law, one first looks at the dispersion law that goes with the field equation (3.37). That is, one assumes that it admits "wave-like" solutions of some specified form and arrives at an algebraic relationship after the differentiations have been performed, which will then define the dispersion law for that type of solution; of course, one must first clarify what one means by the open-ended phrase "wave-like solutions." Here, one must be careful, because unless the field equation is linear (i.e., $K$ is not still a function of $F$), different definitions of wave-like solutions will produce different dispersion laws.

In the linear case, it is sufficient to look at the dispersion law for travelling plane-wave solutions, which will take the general coordinate form:

$$A_\nu = a_\nu e^{i k(\mathbf{x})}, \qquad k(\mathbf{x}) = \omega t - k_i x^i, \qquad (4.12)$$

in which $\omega$ will then represent the frequency of oscillation, and the $k_i$ represent the wave numbers in each spatial direction. In order for one to be dealing with plane-waves, the amplitude $a_\nu$, as well as $\omega, k_i$ must all be constants.

When (4.12) is substituted in (3.37), at any point in space-time outside the support of $\mathbf{J}$ (so $J_\nu = 0$), one will get:

$$\mathcal{D}^{\nu\lambda}(k) a_\lambda \equiv (K^{\mu\nu\kappa\lambda} k_\mu k_\kappa) a_\lambda = 0. \qquad (4.13)$$

In order for this linear equation for $a_\lambda$ to admit non-zero solutions, the determinant of the matrix $\mathcal{D}^{\nu\lambda}(k)$ must vanish, which will yield the dispersion law:

$$0 = \det [\mathcal{D}^{\nu\lambda}(k)] = \det [K^{\mu\nu\kappa\lambda} k_\mu k_\kappa]. \qquad (4.14)$$

In general (e.g., for linear, but anisotropic, media), the right-hand side of this equation will be a homogeneous quartic polynomial in the frequency-wave number 1-form $k$, partly due to the fact that electromagnetic waves are transverse, not longitudinal. For some electromagnetic media, the quartic polynomial will factorize into a product of quadratic polynomials (i.e., bimetricity [**21**]), and in even more restricted cases, it will be the square of a quadratic polynomial. The latter case is the one that gives the Lorentzian metric; i.e., the dispersion law for the classical electromagnetic vacuum or *light-cones.*

In particular, when $K^{\mu\nu\kappa\lambda}$ takes the form (4.7), one will have:

$$\mathcal{D}^{\nu\lambda}(k) = \tfrac{1}{2}[k^2\, g^{\nu\lambda} - k^\nu\, k^\lambda], \quad k^2 \equiv g(k, k). \tag{4.15}$$

That will make:

$$\mathcal{D}^{\nu\lambda}(k)\, a_\lambda = \tfrac{1}{2}[k^2\, a^\lambda - (k^\lambda\, \alpha_\lambda)\, k^\nu]. \tag{4.16}$$

However, with $A$ in the form (4.12), the Lorentz gauge condition (4.10) will become:

$$0 = k_\mu\, a^\mu = k^\mu\, a_\mu, \tag{4.17}$$

which will reduce (4.16) to simply:

$$\mathcal{D}^{\nu\lambda}(k)\, a_\lambda = \tfrac{1}{2} k^2\, a^\lambda, \tag{4.18}$$

and that will vanish iff:

$$k^2 = 0, \tag{4.19}$$

which is the usual Lorentz dispersion law.

When one multiplies the frequency-wave number 1-form $k$ by $\hbar$, one will get the energy-momentum 1-form $p$. In that way, an electromagnetic wave that has the latter dispersion law will be associated with an energy-momentum 1-form that satisfies $p^2 = 0$. Hence, such a wave will carry no mass. One can also show that such a wave will be transverse by assuming that the fields $F$, $\mathfrak{H}$ of the wave have the form of plane waves:

$$F = e^{ik(\mathbf{r})}\hat{F}, \quad \mathfrak{H} = e^{ik(\mathbf{r})}\hat{\mathfrak{H}}, \qquad \mathbf{r} = t\, \partial_t + x^i\, \partial_i, \tag{4.20}$$

in which the components $k_\mu$ of $k = \omega\, dt - k_s$ are constant, as well as $\hat{F}$, $\hat{\mathfrak{H}}$.

For such fields the Maxwell equations in the absence of sources will take the form:

$$k \wedge \hat{F} = 0, \qquad i_k \hat{\mathfrak{H}} = 0 \qquad (k_{[\lambda} \wedge \hat{F}_{\mu\nu]} = 0, \quad k_\mu \hat{\mathfrak{H}}^{\mu\nu} = 0). \tag{4.21}$$

If $\hat{F}$ and $\hat{\mathfrak{H}}$ have the 1+3 forms:

$$\hat{F} = c\, dt \wedge \hat{E} - \hat{B}, \qquad \hat{\mathfrak{H}} = \frac{1}{c}\partial_t \wedge \hat{\mathbf{D}} + \hat{\mathbf{H}}, \tag{4.22}$$

resp., then when they are substituted in equations (4.21) and the independent components are set to zero, one will get:

$$\omega \hat{B} = c\, k_s \wedge \hat{E}, \qquad k_s(\hat{\mathbf{B}}) = 0, \qquad \omega\, \#_s \hat{\mathbf{D}} = -c\, k_s \wedge \hat{H}, \qquad k_s(\hat{\mathbf{D}}) = 0, \quad (4.23)$$

in which we have also defined the spatial vector $\hat{\mathbf{B}}$ and the covector $\hat{H}$ to make:

$$\hat{B} = \#_s(\hat{\mathbf{B}}), \quad \hat{H} = \#_s(\hat{\mathbf{H}}). \quad (4.24)$$

The first equation in (4.23) says that the spatial 2-form $\hat{B}$ defines a spatial plane that is spanned by $k_s$ and $\hat{E}$ in each cotangent space to space-time. One can also make a 1-form $\hat{B}_n$ out of $\hat{B}$ that is basically the normal to the plane of $k_s$ and $\hat{E}$ by first going to the vector $\hat{\mathbf{B}}$ using $\#_s^{-1}$ and then going to the covector $\hat{B}_n$ using the spatial metric $-\delta_{ij}$; i.e.:

$$\hat{B}_i = -\tfrac{1}{2}\, \delta_{ij}\, \varepsilon^{jkl}\, \hat{B}^{kl}. \quad (4.25)$$

Hence, the covector $\hat{B}_n$ is orthogonal to both $k_s$ and $\hat{E}$. The second equation in (4.23) says that the spatial vector $\hat{\mathbf{B}}$ is in the plane that is annihilated by $k_s$, so that condition will amount to saying that $k_s$ is orthogonal to $\hat{B}_n$. In order to show that $\hat{E}$ is orthogonal to $\hat{B}_n$, note that when one takes the exterior products of both sides of the first equation in (4.23) with $\hat{E}$, one will get the vanishing of $\hat{E} \wedge \hat{B}$, but that means that:

$$0 = \hat{E} \wedge \hat{B} = \hat{E} \wedge \#_s \hat{\mathbf{B}} = \hat{E} \wedge i_{\hat{\mathbf{B}}} V_s = -\hat{E}(\hat{\mathbf{B}})\, V_s$$

Hence, the triple $\{\hat{E},\, \hat{B}_n,\, k_s\}$ will be an orthogonal triad of spatial vectors, and if one includes the covector $c\, dt$ then one will get an orthogonal space-time frame $\{c\, dt,\, \hat{E},\, \hat{B}_n,\, k_s\}$ in each cotangent space. Although the covectors $\hat{E}$, $\hat{B}_n$, and $k_s$ are not typically unit covectors, since they are all non-zero, they can be normalized to give orthonormal and Lorentzian frames. In any event, one can say that the "plane of oscillation" – i.e., the one spanned by $\hat{E}$ and $\hat{B}_n$ – is transverse to the propagation covector $k_s$ or that the wave itself is "transverse." The "plane of propagation" is the one that is spanned by $k_s$ and $\hat{E}$.

Similarly, the third equation in (4.23) says that the spatial 2-form $\#_s \hat{\mathbf{D}}$ defines a spatial plane that is spanned by $k_s$ and $\hat{H}$, and the fourth equation says that the covector $k_s$ is orthogonal to $\hat{D}_n$, which is dual to $\hat{\mathbf{D}}$ using the spatial metric; hence, $\hat{D}_n$ is normal to the plane of $\#_s \hat{\mathbf{D}}$. In order to see that $\hat{D}_n$ is orthogonal to $\hat{H}$, start with the fact that:

$$\hat{H}(\hat{\mathbf{D}})\, V_s = -\hat{H} \wedge \#_s \hat{\mathbf{D}}_s$$

and use the third equation in (4.23) . Hence, the triple of covectors { $\hat{D}_n$, $\hat{H}$, $k_s$ } is also orthogonal, as is the tetrad {$c\,dt$, $\hat{D}_n$, $\hat{H}$, $k_s$ }, and both can be normalized to an orthonormal triad and a Lorentzian tetrad, respectively. That means that the wave in question is also transverse by this definition.

An important property of the 2-form $\hat{F}$ is that when one solves the first of (4.23) for $\hat{B}$ and substitutes that into the first of (4.22), one will get:

$$\hat{F} = \frac{1}{\omega c} k \wedge \hat{E}, \qquad (4.26)$$

which says that $\hat{F}$ defines a plane in each cotangent space that is spanned by $k$ and $\hat{E}$, and can also be called the plane of propagation in the four-dimensional sense. As a consequence, one will have:

$$0 = \hat{F} \wedge \hat{F} = -2c^2\, dt \wedge \hat{E} \wedge \hat{B} = 2c^2 \hat{E}(\hat{\mathbf{B}})\, V . \qquad (4.27)$$

Hence, $\hat{F} \wedge \hat{F}$ will vanish iff $\hat{E}(\hat{\mathbf{B}}) = \hat{E}_i \hat{B}^i$ does. Therefore, $\hat{F}$ will define a plane iff $\hat{E}$ is orthogonal to $\hat{B}_n$.

Similarly,

$$\hat{\mathfrak{H}} \wedge \hat{\mathfrak{H}} = \frac{2}{c^2} \partial_t \wedge \hat{\mathbf{D}} \wedge \hat{\mathbf{H}} = -\frac{2}{c^2} \hat{H}(\hat{\mathbf{D}}) V = 0, \qquad (4.28)$$

so $\hat{\mathfrak{H}} \wedge \hat{\mathfrak{H}}$ will vanish iff $\hat{H}(\hat{\mathbf{D}}) = \hat{H}_i \hat{D}^i$ does, and therefore $\hat{\mathfrak{H}}$ will also define a plane of propagation. In fact, if one solves the third equation in (4.23) for $\hat{\mathbf{H}}$:

$$\hat{\mathbf{H}} = \frac{1}{c\omega} \mathbf{k}_s \wedge \hat{\mathbf{D}} \qquad (4.29)$$

and substitutes that in the second of (4.22) then one will get:

$$\hat{\mathfrak{H}} = \frac{1}{\omega c} \mathbf{k} \wedge \hat{\mathbf{D}}, \qquad (4.30)$$

so the plane of $\hat{\mathfrak{H}}$ will be spanned by $\mathbf{k}$ and $\hat{\mathbf{D}}$.

The way that the two spatial frames { $\hat{E}$, $\hat{B}_n$, $k_s$ } and { $\hat{D}_n$, $\hat{H}$, $k_s$ } relate to each other depends upon the constitutive law of the medium. In the simplest case of the classical vacuum, the covector $\hat{D}$ will be collinear with $\hat{E}$, and $\hat{H}$ will be collinear with $\hat{B}_n$. For more general linear algebraic relationships, the relationship between coframes might involve shearing deformations of the frame, and quite often the constitutive law itself becomes a function of frequency.

*c. Five-dimensional waves.* – Although there are various ways of physically interpreting the added fifth dimension to space-time, we shall use the dispersion law for electromagnetic waves in Minkowski space as the basis for our interpretation and the dispersion law for matter waves in quantum mechanics as the basis for its extension.

The dispersion law for massless electromagnetic waves in the classical vacuum (4.19) can be expressed in the form:

$$0 = k^2 = \eta^{\mu\nu} k_\mu k_\nu = \left(\frac{\omega}{c}\right)^2 - (k_1)^2 - (k_2)^2 - (k_3)^2. \tag{4.31}$$

In order to get the corresponding one for massive matter waves, one starts with the relativistic expression for the (rest) mass hyperboloid of point-like mass whose rest mass is $m_0$ and whose energy-momentum 1-form is:

$$p = (E/c)\, dx^0 - p_i\, dx^i = E\, dt - p_i\, dx^i, \tag{4.32}$$

in which $E$ is the total energy of the moving point, and $p$ is its spatial momentum when the curve is parameterized by proper time. The rest mass hyperboloid is then defined by:

$$(m_0 c)^2 = p^2 = \eta^{\mu\nu} p_\mu p_\nu = \left(\frac{E}{c}\right)^2 - (p_1)^2 - (p_2)^2 - (p_3)^2. \tag{4.33}$$

One then introduces the de Broglie relations between $p$ and $k$:

$$p = \hbar k, \qquad E = \hbar \omega, \qquad p_i = \hbar k_i, \tag{4.34}$$

which will convert (4.33) into:

$$k_0^2 = k^2, \tag{4.35}$$

in which the rest mass $m_0$ has been replaced with the corresponding Compton wave number:

$$k_0 = \frac{m_0 c}{\hbar}. \tag{4.36}$$

In particular, $k_0$ will vanish iff $m_0$ does, and one can regard the dispersion law (4.31) as a special case of (4.35). However, the general dispersion law (4.35) will no longer be homogeneous, but if one wishes to make it homogeneous again then one needs only to make a component of $k_0$ itself. One then defines:

$$k_4 = k_0 \tag{4.37}$$

and introduces new indices for the five components $k_A$ with $A = 0, \ldots, 4$.

With that substitution, (4.35) will become:

$$0 = \eta^{\mu\nu} k_\mu k_\nu - (k_4)^2. \tag{4.38}$$

This suggests that we can extend the four-dimensional Minkowski space metric $\eta^{\mu\nu}$ to a five-dimensional one $\eta^{AB}$ by setting the missing components equal to:

$$\eta^{\mu 4} = \eta^{4\mu} = 0, \quad \eta^{44} = -1. \tag{4.39}$$

The general dispersion law for waves, whether massive or massless, will then take the form:

$$0 = k^2 = \eta^{AB} k_A k_B. \tag{4.40}$$

Hence, both kinds of waves – viz., massive and massless – will lie on the five-dimensional "light-cone." Indeed, one can go from the mass hyperboloid to the light-cone by means of a continuous one-parameter family of quadrics by simply letting $k_0$ go to zero.

It is straightforward to define a corresponding five-dimensional d'Alembertian operator:

$$^5\Box = \eta^{AB} \partial_{AB} = \frac{1}{c^2} \frac{\partial^2}{\partial t^2} - \frac{\partial^2}{\partial (x^1)^2} - \frac{\partial^2}{\partial (x^2)^2} - \frac{\partial^2}{\partial (x^3)^2} - \frac{\partial^2}{\partial (x^4)^2}. \tag{4.41}$$

The five-dimensional electromagnetic wave equations that would correspond to the classical four-dimensional vacuum case (4.11) would then be:

$$^5\Box A^C = -4\pi J^C \qquad (C = 0, \ldots, 4). \tag{4.42}$$

A first guess at the five-dimensional constitutive law that produces this would be to use the linear isomorphism that represents the raising of both indices of the 2-form $^5F$, which would have the component form:

$$K^{ABCD} = \eta^{AR} \eta^{BS} - \eta^{AS} \eta^{BR}. \tag{4.43}$$

Under a 4+1 split, this will give the matrices above:

$$K^{\mu\nu\kappa\lambda} = \text{(corresponding components of } K^{ABCD}\text{)}, \tag{4.44}$$

$$L^{\mu\nu\kappa} = K^{\mu\nu\kappa 4} = \eta^{\mu\kappa} \eta^{\nu 4} - \eta^{\mu 4} \eta^{\nu\kappa} = 0, \tag{4.45}$$

$$M^{\mu\kappa\lambda} = K^{\mu 4\kappa\lambda} = \eta^{\mu\kappa} \eta^{4\lambda} - \eta^{\mu\lambda} \eta^{4\kappa} = 0, \tag{4.46}$$

$$N^{\mu\kappa} = K^{\mu 4\kappa 4} = \eta^{\mu\kappa} \eta^{44} - \eta^{\mu 4} \eta^{4\kappa} = -\eta^{\mu\kappa}, \tag{4.47}$$

since all matrices of the form $\eta^{\mu 4}$ must vanish.

That allows us to rewrite the constitutive law in the simpler form:

$$\mathfrak{H}^{\mu\nu} = K^{\mu\nu\kappa\lambda} F_{\kappa\lambda} = F^{\mu\nu}, \qquad h^\mu = -\eta^{\mu\nu} f_\nu = -f^\mu, \tag{4.48}$$

and the second set of Maxwell equations will take on the form:

$$\partial_\mu F^{\mu\nu} = -4\pi J^\nu - \partial_4 f^\nu, \qquad \partial_\mu f^\mu = 4\pi J^4. \tag{4.49}$$

Hence, even in four-dimensional space-time, the Maxwell equations have picked up a contribution from the fifth dimension. Note that the second equation essentially decouples $f^\mu$ from $F^{\mu\nu}$ and $J^\nu$ by making $f$ a field that is determined solely by $J^4$. Hence, the only unknowns in the first equation will be the $F^{\mu\nu}$.

For the sake of clarity, we combine the last set of equations with the first Maxwell equations (2.21):

$$\boxed{d_\wedge F = 0, \quad \partial_4 F = -d_\wedge f, \quad \text{div } \mathbf{F} = -4\pi \mathbf{J} - \partial_4 \mathbf{f}, \quad \text{div } \mathbf{f} = 4\pi J^4.} \tag{4.50}$$

In these equations, we have defined $\mathbf{F}$ to be the bivector field $\frac{1}{2} F^{\mu\nu} \partial_\mu \wedge \partial_\nu$ and $\mathbf{f}$ to be the vector field $f^\mu \partial_\mu$.

When equations (4.50) are expressed in terms of $A_\mu$, $A_4$, and one includes the five-dimensional Lorentz condition, they will take the form:

$$\Box A^\mu - \partial_4^2 A^\mu = -4\pi J^\mu, \qquad {}^5\Box A^4 = 4\pi J^4, \qquad \partial_\mu A^\mu = -\partial_4 A^4, \tag{4.51}$$

which is basically (4.42), along with the gauge condition. Note that the only coupling between the space-time components $A^\mu$ of $A$ and fifth component $A^4$ is by way of the gauge condition.

Assume that $F$, $\mathbf{F}$, and $\mathbf{f}$ take the form of five-dimensional plane waves with a frequency-wave number 1-form ${}^5k = k - k_0\, dx^4$ whose components are constants, so the phase function takes the form:

$$\theta(x^\mu, x^4) = k(\mathbf{r}) - k_0 x^4, \tag{4.52}$$

and:

$$F = e^{i\theta} \hat{F}, \qquad \mathbf{F} = e^{i\theta} \hat{\mathbf{F}}, \qquad f = e^{i\theta} \hat{f}, \qquad \mathbf{f} = e^{i\theta} \hat{\mathbf{f}}. \tag{4.53}$$

In the absence of sources, equations (4.50) will then take the form:

$$k \wedge \hat{F} = 0, \qquad k_0 \hat{F} = k \wedge \hat{f}, \qquad i_\mathbf{k} \hat{\mathbf{F}} = k_0 \hat{\mathbf{f}}, \qquad k(\hat{\mathbf{f}}) = 0. \tag{4.54}$$

The second of these equations allows us to solve for $\hat{F}$ in terms of $k$ and $\hat{f}$, while the last two allow us to solve for $\hat{\mathbf{F}}$ in terms of $\mathbf{k}$ and $\hat{\mathbf{f}}$, namely:

$$\hat{F} = \frac{1}{k_0} k \wedge \hat{f}, \qquad \hat{\mathbf{F}} = \frac{1}{k_0} \mathbf{k} \wedge \hat{\mathbf{f}}. \tag{4.55}$$

That allows us to represent ${}^5\hat{F}$ and ${}^5\hat{\mathbf{F}}$ in the forms:

$$^5\hat{F} = \frac{1}{k_0}\,^5k \wedge \hat{f} = \hat{F} + \hat{f} \wedge dx^4, \qquad ^5\hat{\mathbf{F}} = \frac{1}{k_0}\,^5\mathbf{k} \wedge \hat{\mathbf{f}} = \hat{\mathbf{F}} - \hat{\mathbf{f}} \wedge \partial_4, \tag{4.56}$$

respectively, with:

$$\hat{F} = \frac{1}{k_0} k \wedge \hat{f}, \qquad \hat{\mathbf{F}} = \frac{1}{k_0} \mathbf{k} \wedge \hat{\mathbf{f}}. \tag{4.57}$$

In particular, we will have:

$$^5\hat{F} \wedge ^5\hat{F} = 0, \qquad ^5\hat{\mathbf{F}} \wedge ^5\hat{\mathbf{F}} = 0. \tag{4.58}$$

When $\hat{F}$ is expressed in $\hat{E} - \hat{B}$ form as in (4.22) and $\hat{\mathbf{F}}$, $\hat{f}$, and $\hat{\mathbf{f}}$ are expressed in the form:

$$\hat{\mathbf{F}} = \frac{1}{c}\partial_t \wedge \hat{\mathbf{D}} + \hat{\mathbf{H}}, \qquad \hat{f} = \hat{f}_t\, dt - \hat{f}_s, \qquad \hat{\mathbf{f}} = \hat{f}^t \partial_t + \hat{\mathbf{f}}_s, \tag{4.59}$$

equations (4.54) will initially yield:

$$\omega \hat{B} = c\, k_s \wedge \hat{E}, \quad k_s(\hat{\mathbf{B}}) = 0, \quad c k_0\, \hat{E} = f_t\, k_s - \omega\, \hat{f}_s, \quad k_0\, \hat{B} = -k_s \wedge \hat{f}_s, \tag{4.60}$$

$$\omega \#_s \hat{\mathbf{D}} = -c\,[k_s \wedge \hat{H} + k_0\, \#_s(\hat{\mathbf{f}}_s)], \quad k_s(\hat{\mathbf{D}}) = -c k_0\, \hat{f}_t, \quad k_s(\hat{\mathbf{f}}_s) = \omega \hat{f}^t. \tag{4.61}$$

However, these equations are not all independent. In particular, since the third and fourth equations in (4.60) allow one to define $\hat{E}$ and $\hat{B}$ in terms of $k$ and $\hat{f}$, the first two equations will become identities. One can then say that:

$$\hat{F} = \frac{1}{k_0}(f_t\, dt \wedge k_s - k \wedge \hat{f}_s), \qquad \hat{\mathbf{F}} = \frac{1}{k_0}(f^t \partial_t \wedge \mathbf{k}_s + \mathbf{k} \wedge \hat{\mathbf{f}}_s). \tag{4.62}$$

Hence, we are beginning to see that the spatial 1-form $\hat{f}_s$ seems to play a role that is similar to that of $\hat{E}$, while the presence of $f_t$ has altered the orientation of the plane of propagation from the plane of $k$ and $\hat{f}_s$. Similarly, the role of $\hat{\mathbf{f}}_s$ is analogous to that of $\hat{\mathbf{D}}$.

Although $\hat{f}$ seems to play the fundamental role now, it is still useful to know that its introduction has not changed the basic relationships between $\hat{E}$, $\hat{\mathbf{B}}$, and $k_s$. Note that the 3-form $\hat{E} \wedge \hat{B}$, which is proportional to $\hat{E}(\hat{\mathbf{B}})$, will still vanish.

Meanwhile, what has changed are the relationships between $\hat{\mathbf{D}}$, $\hat{H}$, and $k_s$. In particular, the first equation in (4.61) says that the plane of the 2-form $\#_s \hat{\mathbf{D}}$ is no longer spanned by $k_s$ and $\hat{H}$, while the second one says that the spatial covector $\hat{D}_n$ that is metric dual to the vector $\hat{\mathbf{D}}$ is no longer orthogonal to $k_s$. That has the effect of introducing a "longitudinal" component to $\hat{D}_n$ that is collinear with $k_s$. Hence, a massive electromagnetic wave will typically not be transverse.

**5. Quantum wave equations.** – In order to get a better idea of how the added space-time dimension has enlarged the scope of Maxwell's equations, it helps to look at how they include the wave equations of relativistic quantum mechanics.

*a. Klein-Gordon equation.* – The Klein-Gordon equation for a complex-valued wave function $\Psi$, namely:

$$\Box \Psi + k_0^2 \Psi = 0, \tag{5.1}$$

can be obtained from the five-dimensional wave equation:

$$^5\Box\, ^5\Psi = 0 \tag{5.2}$$

by separating the fifth variable $x^4$ in a manner that is analogous to the way that one obtains the Helmholtz equation from the four-dimensional one by separating the time variable. Hence, one assumes that $^5\Psi$ is the product of a function of $x^4$ with a function of the space-time coordinates $x^\mu$:

$$^5\Psi (x^\mu, x^4) = \Psi (x^\mu)\, \Phi (x^4). \tag{5.3}$$

When that is substituted in (5.2), it will take the form:

$$\frac{1}{\Psi} \Box \Psi (x^\mu) = \frac{1}{\Phi} \frac{d^2}{d(x^4)^2} \Phi (x^4),$$

and since the two sides of the equation are functionally independent of each other, they can only be equal to a constant. If we call that separation constant $-k_0^2$ then the five-dimensional wave equation (5.2) will separate into the Klein-Gordon equation for $\Psi$ and a simple harmonic oscillator equation for $\Phi$:

$$\frac{d^2 \Phi}{d(x^4)^2} + k_0^2 \Phi = 0. \tag{5.4}$$

The latter equation admits solutions that are linear combinations of sinusoidal functions of $x^4$ that can be expressed in the form $e^{\pm i k_0 x^4}$.

That same process can be applied to any other solutions of five-dimensional wave equations, such as the five-dimensional electromagnetic potential 1-form $^5A$ at points outside the support of $^5\mathbf{J}$. If we assume that its components $^5A_\mu$ – or rather $^5A^\mu$ – are products of the form:

$$^5A^\mu (x^\nu, x^4) = \mathfrak{a}^\mu (x^\nu)\, \Phi (x^4) \tag{5.5}$$

and call the separation constant $k_0^2$, as before, then the five-dimensional wave equation for $^5A^\mu$ will separate into a Klein-Gordon equation for each $\mathfrak{a}^\mu(x^\nu)$:

$$\Box\, \mathfrak{a}^\mu + k_0^2\, \mathfrak{a}^\mu = 0 \tag{5.6}$$

and (5.4), which will make $\Phi$ into an exponential phase factor $e^{\pm i k_0 x^4}$ that depends upon $x^4$.

Equation (5.6) is called the *Proca equation* [**22, 23**]. It can also be obtained from the first equation in (4.51) by separating the variables in $A^\mu$ as in (5.5). Proca was originally looking for a relativistic equation for the electron that would represent an alternative to the Dirac equation.

If the $\mathfrak{a}^\mu$ are themselves exponential – viz., plane waves – then the dispersion law for such waves will become:

$$k^2 = k_0^2. \tag{5.7}$$

Of course, this is the usual dispersion law for massive matter waves in quantum mechanics, and the traditional objection to the use of massive electromagnetic waves in four dimensions was that if one were to give mass to the photon then it would also have to exhibit longitudinal modes of vibration, in addition to the better-established transverse ones. However, we have now essentially embedded four-dimensional wave motion into the larger context of the propagation of *five*-dimensional electromagnetic waves. That is: The five-dimensional formulation admits the possibility of *both* massive *and* massless electromagnetic waves, as well as a possible "electromagnetic" interpretation of the quantum wave function.

*b. Dirac equation.* – The usual limitation to the Klein-Gordon equation that early relativistic quantum theory was addressing was that it did not account for the possibility that the wave function described a particle with spin, such as the electron. The first difference that one encounters in attempting to include spin is that the field space (viz., the vector space in which the wave function $\Psi$ takes its values) must be extended from the one-complex-dimensional space $\mathbb{C}$ to a two or four-complex dimensional vector space, namely, $\mathbb{C}^2$ or $\mathbb{C}^4$, resp. The former space had been used by Pauli [**24**] in order to include spin in the Schrödinger equation, so that theory was manifestly non-relativistic.

Although a relativistic Pauli equation [**25**] eventually emerged that amounted to an extension of the Klein-Gordon equation, the first attempt to include spin in a relativistic wave equation was Dirac's quantum theory of the electron in 1928 [**26**]. The first step in obtaining his wave equation for the electron (and any other fermion) was to look for a "square root" of the d'Alembertian operator, which would be a first-order linear differential operator $\not{\partial} \equiv \gamma^\mu \partial_\mu$ such that:

$$\Box = \eta^{\mu\nu} \partial_\mu \partial_\nu = \not{\partial}^2 = (\gamma^\mu \partial_\mu)(\gamma^\nu \partial_\nu) = \gamma^\mu \gamma^\nu \partial_\mu \partial_\nu = \tfrac{1}{2}(\gamma^\mu \gamma^\nu + \gamma^\mu \gamma^\nu)\, \partial_\mu \partial_\nu,$$

in which the last step follows from the symmetry of mixed second partial derivatives.

Hence, in order for the last relation to be consistent, the coefficients $\gamma^\mu$ would have to belong to an algebra with the property that:

$$\gamma^\mu \gamma^\nu + \gamma^\mu \gamma^\nu = 2\eta^{\mu\nu}, \tag{5.8}$$

which is, of course, the defining condition for the generators of a Clifford algebra. In particular, it would be the Clifford algebra over four-dimensional Minkowski space.

One can define a Clifford algebra over any orthogonal space, and in particular, five-dimensional Minkowski space, which is what we seem to be dealing with. The straightforward extension of the calculation above to the five-dimensional d'Alembertian $^5\Box$ as the square of an operator of the form $^5\partial\!\!\!/ \equiv \gamma^A \partial_A$ ($A = 0, \ldots, 4$) would imply that the five coefficients $\gamma^A$ would now have to satisfy the condition:

$$\gamma^A \gamma^B + \gamma^B \gamma^A = 2\eta^{AB}. \tag{5.9}$$

If one introduces a 4+1 split on the underlying vector space then these conditions can be also written in the form:

$$\gamma^\mu \gamma^\nu + \gamma^\mu \gamma^\nu = 2\eta^{\mu\nu}, \qquad \gamma^\mu \gamma^4 + \gamma^4 \gamma^\nu = 0, \qquad (\gamma^4)^2 = -1. \tag{5.10}$$

The $\gamma^A$ then represent simply the generators of the Clifford algebra of five-dimensional (real) Minkowski space. Since the dimension of a Clifford algebra over any $n$-dimensional orthogonal space is $2^n$, the algebra that is generated by the $\gamma^A$ will have a (real) dimension of 32, which is twice the dimension of the Clifford algebra $\mathcal{C}(4, \eta)$ over four-dimensional (real) Minkowski space.

An interesting aspect of the dimension number 32 is that although any Clifford algebra admits an isomorphic representation in a matrix algebra, the usual representations of $\mathcal{C}(4, \eta)$ by 4×4 complex matrices (the gamma matrices) are not isomorphic, since the real dimension of $\mathcal{C}(4, \eta)$ is 16, while that is the *complex* dimension of the algebra $M(4, \mathbb{C})$; hence, its real dimension is 32. Hence, the matrix Clifford algebra that is generated by the four gamma matrices is only "half" of the full matrix algebra $M(4, \mathbb{C})$. However, the four gamma matrices do generate an isomorphic representation of the complex Clifford algebra of complex four-dimensional Minkowski space.

Based on dimensional considerations, one might expect that there would be some isomorphic representation of $\mathcal{C}(5, \eta)$ by 4×4 complex matrices. Since the four generators $\gamma^\mu$ already define an isomorphic representation of $\mathcal{C}(4, \eta)$ in a subspace of $M(4, \mathbb{C})$, presumably all that might be required is to represent $\gamma^4$ as a 4×4 complex matrix in order to generate the remaining matrices in $M(4, \mathbb{C})$ by multiplying all of the 16 linearly-independent products of the $\gamma^\mu$ by $\gamma^4$. That would then produce 16 linearly-independent basis vectors that would span a subspace that would be complementary to the subspace that is generated by the $\gamma^\mu$.

The last condition in (5.10) suggests that multiplication by $\gamma^4$ might behave like multiplication by $i$. However, multiplication by $i$ would *commute* with all $\gamma^\mu$, whereas the second condition says that $\gamma^4$ must *anti*-commute with all $\gamma^\mu$. Thus, the matrix $i\,I$ would not be a proper representation of $\gamma^4$. Pauli [10] extended the four generators $\gamma^\mu$ by the product $\gamma^4 = \gamma^0\gamma^1\gamma^2\gamma^3$, which does satisfy the properties that $(\gamma^4)^2 = -1$ and $\gamma^4\gamma^\mu = -\gamma^\mu\gamma^4$. However, one can see that although that element of the algebra is linearly-independent of the other generators, it is not algebraically-independent, since it is defined by a product of the other generators. Hence, that would not be a suitable candidate for $\gamma^4$, either.

One finds that choosing:

$$\gamma^4 = i \begin{bmatrix} 0 & I \\ \hline I & 0 \end{bmatrix} \tag{5.11}$$

will work for both the original Dirac representation of the $\gamma^\mu$, namely:

$$\gamma^0 = \begin{bmatrix} I & 0 \\ \hline 0 & -I \end{bmatrix}, \qquad \gamma^i = \begin{bmatrix} 0 & \sigma^i \\ \hline -\sigma^i & 0 \end{bmatrix}, \tag{5.12}$$

as well as the Weyl representation:

$$\gamma^0 = \begin{bmatrix} 0 & I \\ \hline -I & 0 \end{bmatrix}, \qquad \gamma^i = \begin{bmatrix} 0 & -\sigma^i \\ \hline \sigma^i & 0 \end{bmatrix}. \tag{5.13}$$

The representation of the five-dimensional d'Alembertian as the square of the five-dimensional Dirac operator means that the five-dimensional wave equation will give way to:

$$0 = {}^5\!\partial\!\!\!/\, {}^5\Psi = \gamma^A \partial_A \, {}^5\Psi . \tag{5.14}$$

Although this takes the form of the massless Dirac equation, in the present case we can use rest mass $m_0$ (or more precisely, its Compton wave number $k_0$) as a separation constant and look for wave functions of the form:

$${}^5\Psi (x^\mu, x^4) = e^{ik_0 x^4} \Psi(x^\mu) . \tag{5.15}$$

Substitution of that form for ${}^5\Psi$ in (5.14) will initially yield:

$$0 = e^{ik_0 x^4}(\partial\!\!\!/\,\Psi + ik_0\Psi),$$

which will then give the usual four-dimensional Dirac equation for $\Psi$:

$$\partial\!\!\!/\,\Psi + ik_0 \Psi = 0 \, . \tag{5.16}$$

**6. Five-dimensional relativity.** – Since we have obtained our extension of Minkowski space from four dimensions to five on the basis of the dispersion law for electromagnetic waves in five-dimensional space, it would be good to examine how that would affect the other things that one considers in special relativity. In particular, we need to interpret the extension of the space-time coordinates $x^\mu$ by $x^4$.

The usual definition of the proper time curve parameter $\tau$ is that the 4-velocity vector field:

$$\mathbf{u}\,(\tau) \equiv \frac{dx}{d\tau} = u^\mu\,(\tau)\,\partial_\mu \tag{5.17}$$

of a curve $x\,(\tau) = (x^0\,(\tau), \ldots, x^0\,(\tau))$ in Minkowski space must satisfy:

$$\eta\,(\mathbf{u}, \mathbf{u}) = \eta_{\mu\nu}\, u^\mu u^\nu = (u^0)^2 - (u^1)^2 - (u^2)^2 - (u^3)^2 = c^2. \tag{5.18}$$

Hence, it is tempting to introduce a fifth component $u^4$ to $\mathbf{u}$ that will make:

$$u^4 \equiv \frac{dx^4}{d\tau} = c. \tag{5.19}$$

One can then express (5.18) in homogeneous form:

$$0 = {}^5\eta\,({}^5\mathbf{u}, {}^5\mathbf{u}) = \eta_{AB}\, u^A u^B = (u^0)^2 - (u^1)^2 - (u^2)^2 - (u^3)^2 - (u^4)^2. \tag{5.20}$$

(5.19) can then be integrated to give the definition of $x^4$:

$$x^4 = c\,\tau. \tag{5.21}$$

One should immediately observe that there are deep fundamental issues associated with the difference between curve parameters and local coordinates. In particular, in order to make a coordinate out of proper time, one needs to be dealing with a spatially-distributed congruence of time-like curves, not just a single curve. Furthermore, whether a common parameterization can be given to all of them that would make it possible to compare parameterizations of distinct curves would amount to a question of the integrability of the covelocity 1-form, or rather, the Pfaff equation that it defines. However, one could argue that the same criticism also applies to time, the coordinate $t$. Namely, perhaps what one is calling time, the coordinate, is simply someone else's proper time. This fundamental distinction between time, the parameter, and time, the coordinate, was debated in the early days of relativity, and especially by Max von Laue [**27**], whose lectures on relativity were widely cited for many years ([1]).

---

([1]) One might also confer the author's discussion of proper-time foliations in [**28**].

As a result of the definition (5.21), the form of (5.18) that one obtains by multiplying both sides by $(d\tau)^2$, namely:

$$c^2 (d\tau)^2 = \eta_{\mu\nu} dx^\mu dx^\nu, \tag{5.22}$$

can also be expressed in homogeneous form:

$$\eta_{AB} dx^A dx^B = 0. \tag{5.23}$$

As pointed out before, the extension of the frequency-wave number 1-form $k = k_\mu dx^\mu$ by the fifth component $k_4 = m_0 c / \hbar$ is consistent with the extension of the energy-momentum 1-form $p = p_\mu dx^\mu$ by the fifth component $p_4 = m_0 c$. Finally, if we assume that the electric current four-vector $\mathbf{J}$ is spatially-distributed with a charge density $\sigma_0$ in the rest space then if one assumes that the current is "convective" in character, so $\mathbf{J} = \sigma_0 \mathbf{u}$, then the obvious extension of that definition to five dimensions would be:

$$^5\mathbf{J} = \sigma_0 \,{}^5\mathbf{u}, \tag{5.24}$$

which would make:

$$J^4 = \sigma_0 u^4 = \sigma_0 c. \tag{5.25}$$

That definition will allow us to refine the field equations (3.22) to now read:

$$\boxed{\begin{array}{ll} d_\wedge F = 0, & \partial_\tau F = -c\, d_\wedge f, \\ \operatorname{div} \mathfrak{H} - \dfrac{1}{c}\partial_\tau \mathbf{h} = -4\pi \mathbf{J}, & \operatorname{div} \mathbf{h} = -4\pi c\, \sigma_0, \quad \operatorname{div} \mathbf{J} = -\partial_\tau \sigma_0. \end{array}} \tag{5.26}$$

The main advances over the previous version are to make the vector field $\mathbf{h}$ something that is determined solely by the electric charge density in the rest space and to clarify that the departure from conservation of charge would be due to a change in that density along the world-lines of its motion.

It is important to observe that not all electrical currents in classical electromagnetism are convective. Of particular interest are the "polarization" currents that can arise in polarizable media and take the form of vector fields that are the divergences of bivector fields, namely, the polarization bivector field.

We then summarize these extensions collectively:

$$x^4 = c\,\tau, \quad u^4 = c, \quad k_4 = \frac{m_0 c}{\hbar}, \quad p_4 = m_0 c, \quad J^4 = \sigma_0 c. \tag{5.27}$$

Although our first intuition is to think that the extension to five-dimensional Minkowski space implies that the new invariance group should be $SO(4, 1)$, instead of $SO(3, 1)$, nonetheless, on second glance, one must notice that since both massive and massless motion lie on the same five-dimensional light-cone, in effect both types of motion should be invariant under the conformal group that preserves that five-dimensional light cone. That group would then be $CO(4, 1)$; i.e., the conformal five-dimensional Lorentz group.

**7. Discussion.** – The extension of Maxwell's theory of electromagnetism from four to five dimensions that was just presented is, admittedly, somewhat preliminary. In particular, there is considerable room for development as far as the extensions of the various four-dimensional constitutive laws are concerned, such as the laws that pertain to optical media, in particular. There is also much to be done in regard to the physical interpretation of the space-time vector field **h** and the 1-form *f* that were introduced by the dimensional extension, as well as the extension of the constitutive law that would couple them.

However, the facts that the resulting extension allows for both massless and massive electromagnetic waves and that the most elementary dispersion law for five-dimensional waves includes both the classical electromagnetic waves *in vacuo* and the matter waves that were first introduced by de Broglie and developed by Schrödinger, Klein, Pauli, Dirac, and the others is quite suggestive that relationship between quantum mechanics and relativity is more fundamental than axiomatic. Considering the ubiquitous role that is played by vacuum polarization in quantum electrodynamics, one further suspects that such considerations could be advantageously applied to the extensions of the constitutive law for the classical electromagnetic vacuum to the quantum electromagnetic vacuum.

Another potentially-profitable direction to pursue is the role of projective geometry in the aforementioned extension. One starts to get an impression that the asymptotic limit in which *c* goes to infinity, and all physical parameters (divided by *c*) go to their rest values suggests that the Galilean-Newtonian limit of relativistic mechanics has a lot in common with the points at infinity in a projective space. Furthermore, the usual way of converting inhomogeneous polynomials into homogeneous ones is by regarding the affine coordinates that the inhomogeneous polynomial is defined in terms of as inhomogeneous coordinates for a projective space of the same dimension and then introducing homogeneous coordinates that would project to the inhomogeneous ones. The author has already written other articles on the role of projective geometry in relativity (e.g., [**29**]).

____________